\documentclass[10pt]{elsarticle}
\usepackage{epsfig}

\begin{document}

\title{Unruh effect in vacua with anisotropic scaling: Applications to
multilayer graphene}

\author[RUN]{M.I.~Katsnelson}

\author[AU,ITP]{G.E.~Volovik}

\author[ITEP]{M.A.~Zubkov\footnote{Corresponding author,
 e-mail: zubkov@itep.ru}}

\address[RUN]{Radboud University Nijmegen, Institute for Molecules
and Materials, Heyndaalseweg 135, NL-6525AJ Nijmegen, The Netherlands }

\address[AU]{Low Temperature Laboratory, School of Science and
Technology, Aalto University,  P.O. Box 15100, FI-00076 AALTO, Finland}

\address[ITP]{L. D. Landau Institute for Theoretical Physics,
Kosygina 2, 119334 Moscow, Russia}

\address[ITEP]{ITEP, B.Cheremushkinskaya 25, Moscow, 117259, Russia
}

\begin{abstract}
We extend the calculation of the Unruh effect to the universality classes of
quantum vacua obeying topologically protected invariance under anisotropic
scaling ${\bf r} \rightarrow b {\bf r}$ , $t \rightarrow b^z t$.  Two
situations are considered. The first one is related to the accelerated
detector  which detects the electron - hole pairs. The second one is related
to the system in external electric field, when the electron - hole pairs are
created due to the Schwinger process. As distinct from the Unruh effect in
relativistic systems (where $z=1$) the calculated radiation is not thermal,
but has properties of systems in the vicinity of  quantum criticality. The
vacuum obeying  anisotropic scaling can be realized, in particular, in
multilayer graphene with the rhombohedral stacking. Opportunities of the
experimental realization of Unruh effect in this situation are discussed.
\end{abstract}



\maketitle


\section{Introduction}

Relativistic physics can be considered as emergent phenomenon,
which occurs in the low-energy limit in some universality classes of quantum
vacua with topologically protected nodes in the energy spectrum.
The general topological approach to classification of gapless vacua, which
is based on connection to the $K$-theory, has been developed by Ho\v{r}ava
\cite{Horava2005}.   Ho\v{r}ava demonstrated that the reason for the
emergence of Lorentz invariance is the topological theorem, called the
Atiyah-Bott-Shapiro construction.

The Lorentz invariance emerges only for the particular classes of
topological vacua, where nodes in energy spectrum are characterized by the
elementary topological charges, $N=\pm 1$. For the higher values of the
winding number,  the Lorentz invariance can be supported by symmetry between
the fermionic species. This symmetry makes the vacuum to be effectively
equivalent to system of fermions, each having elementary topological charges
$N=\pm 1$, see \cite{Volovik2003}.
If such symmetry is absent but the splitting of the nodes in the energy
spectrum into the elementary nodes is still forbidden, then in the quantum
vacua with higher values of topological number $|N| > 1$ the low-energy
fermionic spectrum obeys the invariance under anisotropic scale
transformation. An example of the anisotropic scale transformation is ${\bf r}
\rightarrow b {\bf r}$ , $t \rightarrow b^z t$ with $z\neq 1$, i.e. one has
different scale transformation for space and time and thus contrary to the
relativistic systems with $z=1$ space and time cannot be united into
space-time. Systems with anisotropic scaling in 3+1 and 2+1 vacua are
discussed in Ref. \cite{Volovik2003}
and in Refs.
\cite{Volovik2007,Manes2007,Dietl-Piechon-Montambaux2008,Banerjee2009,HeikkilaVolovik2010,KatsnelsonVolovikZubkov2012}, respectively. In a simple case
of 2+1 systems, such as multilayered graphene \cite{Katsbook}, the scaling
parameter $z$ coincides with topological charge of the node in the energy
spectrum, $z=|N|$, in other words, the anisotropic scaling is topologically
protected.

Physics obeying the topologically protected anisotropic scaling is
fundamentally different from the relativistic physics. It emerges in the
vacua, which belong to different topological classes, and thus  cannot
overlap with the relativistic physics. The transition between different
topological classes cannot occur adiabatically. That is why the transfer
from the vacuum with emergent relativistic physics to the anisotropic
vacuum with a given scaling parameter $z \neq 1$ occurs only by
discontinuous jump -- by quantum phase transition. The same refers to the
quantum transitions between anisotropic vacua with different $z$. Thus, the
topologically protected quantum vacua with anisotropic scaling open the new
natural area for application of quantum field theory. For example, while
traditional relativistic QED was considered as the function of the space
dimension $D$ \cite{Blau1991}, the general QED must be considered as
function of two parameters, $D$ and $N$
\cite{KatsnelsonVolovikZubkov2012}.
The QED  with the anisotropic scaling can be realized in the multilayered
graphene, and it is essentially richer than the relativistic QED with
$z=1$. The modified  Heisenberg-Euler action  describing the polarization of
the anisotropic quantum vacuum with $z\neq 1$ has been discussed in Refs.
\cite{KatsnelsonVolovikZubkov2012,KatsnelsonVolovik2012,Zubkov2012}. In
particular, this effective action has different scaling laws for electric
and magnetic fields.

The anisotropic scaling  can be useful also to quantum gravity. For example,
the vacua with topologically protected anisotropic scaling can host the
Ho\v{r}ava gravity, which is based just  on the idea of anisotropic scaling
\cite{HoravaPRL2009,HoravaPRD2009,Horava2008}.
The advantage of the Ho\v{r}ava extension of general relativity is that for
$z=3$ the quantum gravity can be  ultraviolet complete.

The natural question is how the other quantum phenomena, such as Hawking
radiation \cite{Hawking1974} and Unruh effect \cite{Unruh1976}, which are
intrinsic to the quantum vacuum,  are modified compared to the relativistic
case.
The related phenomenon -- Schwinger pair creation in electric field in
anisotropic vacuua -- has been considered in Refs.
 \cite{KatsnelsonVolovikZubkov2012,KatsnelsonVolovik2012,Zubkov2012}.
In the relativistic vacuum, the Bekenstein  conjecture of the generalized
second law of the black hole thermodynamics
\cite{Bekenstein1973} and Hawking conjecture of the black hole radiance
\cite{Hawking1974}
indicated the close connection between general relativity (GR) and
thermodynamics of the quantum vacuum.
This was followed by the Unruh conjecture \cite{Unruh1976} that the detector
moving in the Lorentz invariant vacuum with constant acceleration views the
quantum vacuum as a thermal background. The latter indicated the close
connection between the thermodynamics and the vacuum in special relativity.
Then the thermal behavior of the quantum  vacuum has been extended to the
expanding Universe \cite{GibbonsHawking1977}. At the same time, it was found
that the vacuum of Lorentz invariant quantum electrodynamics (QED) in an
exactly constant electric field {\bf E}  can be also understood as a thermal
background characterized by the temperature parameter \cite{Muller1977} (for
recent papers see
\cite{LabunRafelski2012}).

All these phenomena occurring in the relativistic quantum vacuum are
interrelated and have similar properties. In particular, the power spectrum
of the vacuum noise (or the detector-response function) seen by a uniformly
accelerated observer in flat spacetimes depends on space dimension $D$, and
for even $D$ it exhibits the phenomenon of the apparent inversion of
statistics  \cite{Takagi1986}.
A similar phenomenon of the inversion of spin-statistics relation occurs for
the vacuum polarization in
a constant electric field  \cite{Muller1977,PauchyHwang2009}.
As another example, it is argued that the same back reaction effect of
particle creation influences the vacuum of QED in electric field and the
quantum vacuum in the de Sitter space-time  \cite{KrotovPolyakov2011}.
This implies the instability of the de Sitter vacuum towards decay, which
remains the issue of controversy
\cite{StarobinskyYokoyama1994,Polyakov2008,Polyakov2010,Akhmedov2012,Klinkhamer2012}.
Extension of the stability analysis to de Sitter vacuum with different
$z\neq 1$ has been considered in Ref.  \cite{Huang2010}.

Here we discuss the modification of  the thermal properties of the quantum
vacuum as compared to the Lorentz invariant vacua.
We consider the particle production in electric field and the Unruh effect
in the anisotropic scaling system, such as multilayered graphene.
In general anisotropic case the energy distribution of electron-hole pairs
in electric field and the distribution detected by the accelerated observer
do not look as thermal. Nevertheless the distributions depend on the same
dimensionless quantity $E/T$, where $T$ is some effective parameter of
dimension of energy, which is determined by electric field or acceleration
correspondingly and by the integer parameter $z$. This indicates the new
type of quantum criticality
\cite{Witczak-Krempa2012,Witczak-Krempa_Sachdev2012,Faulkner2010,Limelette2013,Sachdev2011,GouterauxKiritsis2011}, emerging in the anisotropic vacua
with $z\neq 1$. We will also discuss the opportunities to study these
effects in the multilayer graphene experimentally.

\section{Unruh temperature in multilayer graphene}

In this section we extend the notion of Unruh temperature $T_U$ to the field
theory of multilayer graphene (with ABC stacking). Namely, we introduce it
as the typical energy of processes that take place in the presence of
external force $F$ acting on the quasiparticles.  (In the  case of external
electric field $\cal E$ we have $F = e {\cal E}$.) We consider classical
motion of quasiparticles, introduce the generalized acceleration that
characterizes this motion, and relate it to the Unruh temperature.  We give
qualitative estimate of the ``typical'' energy in graphene in the presence of
external force. This estimate is considered as the definition of Unruh
temperature and will be given in Eq. (\ref{TU2}). It will be shown in the
following sections that this temperature enters the considered distributions
over energy $\omega$ of the electron   - hole pairs in a number of
situations. Namely these distributions depend on the ratio $\omega/T_U$.

\subsection{Classical motion of quasiparticles}

We deal with the two-component spinors placed in the external electric field
directed along the $x$-axis. We introduce the external force $F$ acting on
the fermion excitation adding the potential energy
$F x $ to the one-particle Hamiltonian. Then the Hamiltonian for low-energy
fermionic quasiparticles propagating in the system with anisotropic scaling
$z=J$ has  the form
\begin{equation}
H =  H_0 + Fx, \quad H_0 =   \left(\begin{array}{cc} 0 & v \Bigl(\hat{p}_x -
i \hat{p}_y\Bigr)^J \\
v \Bigl(\hat{p}_x  + i \hat{p}_y\Bigr)^J & 0
\,.
\label{H1}
\end{array}\right)
\end{equation}
Here $v=v_F$ is the Fermi velocity for the case of monolayer ($J=1$) and
$v=1/2m$ for the case of bilayer ($J=2)$, where $m \approx 0.028 m_e$
\cite{mayorov} and $m_e$ is the free-electron mass. In general case  $v
\propto  t_{\bot}^{1-J}v_F^J $, where  $t_{\bot}$ is the interlayer hopping
parameter.
Here and further in the text we work in the system of units with $\hbar =
1$.
Schr\"{o}dinger equation has the usual form
\begin{equation}
i\partial_t \Psi = H \Psi \,.
\end{equation}
We consider the fermion excitations with fixed value of the transverse
momentum $p_y$: $\Psi(t,x,y) = e^{i p_y y} \psi(t,x) $.
 The conserved transverse momentum   $p_y$  plays the role of mass of the
one-dimensional fermion (multiplied by $v_F$ instead of the speed of light).
We denote $p_y = M = {\cal M}v_F$. The corresponding energy spectrum is
$E=\pm v  \Bigl({p_x}^2 + {M}^2\Bigr)^{J/2} =  \pm t_{\bot}^{1-J}v_F^J
\Bigl({p_x}^2 + {\cal M}^2v_F^2\Bigr)^{J/2} $.
 Since $p_y$ plays the role of mass, it is natural to identify the
generalized acceleration as $a = F/{\cal M}= v_F F/p_y$. To justify this
choice, let us apply classical approximation to $\psi(t,x)$ with the
corresponding classical Hamiltonian
\begin{equation}
H_{cl}(p_x,x) =  Fx + v \Bigl({p_x}^2 + {p}^2_y\Bigr)^{J/2} \,.
\label{ClassicalHamiltonian}
\end{equation}
Hamilton equations of motion give classical trajectory that corresponds to
$\dot{x}(0)=0$:
\begin{equation}
x(t) = v F^{J-1} \Bigl({t}^2 + \frac{v_F^2}{a^2}\Bigr)^{J/2} - v  F^{J-1}
\frac{v_F^J}{a^J} + x(0) ~~,~~ a = \frac{F}{\cal M}= v_F\frac{F}{p_y}\,.
\label{traj}
\end{equation}
This is the generalization of the relativistic $J=1$ case, where
Eq.(\ref{traj}) describes the hyperbolic motion with correctly determined
linear acceleration  $a$.
The particular case $p_y=0$ corresponds to the trajectory
\begin{equation}
x(t) = v F^{J-1} |t|^J + x(0)
\label{traj0}
\end{equation}

\subsection{Unruh temperature}

As it was mentioned above we define the Unruh temperature $T_U$ as the
``typical'' energy of processes taking place in the system under the action
of the external force. We estimate this energy semiclassically using the
above mentioned classical trajectory for the particular case $p_y=0$.
Hamiltonian equations relate classical momentum to time as $p_{cl} = F t$.
In Eq. (\ref{ClassicalHamiltonian}) both terms of the Hamiltonian are
typically of the same order. So, we can estimate the energy scale as $T_U
\sim v p_{cl}^J \sim v t^J F^J$. Here typical values of time are related to
energy by the uncertainty
relation $t T_U \sim 1$. Thus we get
\begin{equation}
T_U \sim \Bigl(v F^J\Bigr)^{\frac{1}{J+1}}
\end{equation}
For $J=1$ we shall typically use the definition $T_U =  \Bigl(v_F
F\Bigr)^{\frac{1}{2}}/(2\pi)$. For $J\ge 2$ we shall define $T_U = \Bigl(v
F^J\Bigr)^{\frac{1}{J+1}}$. For the completeness let us rewrite the
expression for the Unruh temperature at $J\ge 2$ in the usual system of
units (with the dimensional Fermi velocity for the single-layer graphene
$v_F$ and the Plank constant $\hbar$ restored):
\begin{equation}
T_U =  t_{\bot} \Bigl(\frac{v_F F \hbar}{t^2_{\bot}}\Bigr)^{\frac{J}{J+1}}
\label{TU2}
\end{equation}
with $t_{\bot } \approx 0.4$ eV, and $v_F \approx c/300$, where $c$ is the
speed of light. Here it is used that $v =  t_{\bot}^{1-J}v_F^J $.

\subsection{Generalized acceleration}

The  identification of Eq. (\ref{traj}) for the generalized acceleration
does not work for the case of Eq. (\ref{traj0}), i.e. for $p_y=0$. This is
the case that was used above for the estimate of the ``typical'' energy
$T_U$.
Unruh temperature in (quasi) relativistic models is related to the
acceleration as $T_U =  a/(2\pi v_F)$. This allows us to estimate the
``typical'' acceleration in the presence of the external force $F$ for the
multilayer graphene.  For $J\ge 1$ we define the acceleration as equal to
the Unruh temperature of Eq. (\ref{TU2}) multiplied by the Fermi velocity
$v_F$.

 The same result is obtained when we define the generalized acceleration as
$F/M_{eff}$, where the effective mass $M_{\rm eff}$ is played by vacuum
fluctuations of the $p_x$ projection of momentum,  $\langle |p_x| \rangle$.
The latter  can be found from the Hamiltonian
(\ref{ClassicalHamiltonian}) at $p_y=0$.
The contribution of
two terms  to the energy of vacuum fluctuations is of the same order,
while the fluctuations $\langle |x|\rangle $ and  $\langle |p_x|\rangle $
are related by the Heisenberg
uncertainty relation: $\langle |x|\rangle \sim  1 / \langle|p_x|\rangle$.
Then one has $F/\langle |p_x|\rangle  = v \langle |p_x|\rangle^J$
which gives  $M_{\rm eff}\sim \langle|p_x|\rangle/v_F = (F/v)^{1/(J+1)}/v_F$
and the generalized acceleration
$a=F/M_{\rm eff}$.

In the special case $J=2$ the equation $a=F/M_{\rm eff}$ is applicable also
for massive case $M\neq 0$. The reason for that is that  the trajectory
(\ref{traj}) for $J=2$
\begin{equation}
x(t) =vF {t}^2 + x(0) = \frac{F {t}^2}{2m} + x(0)   \,,
\label{traj2}
\end{equation}
does not depend on $M$ and thus the generalized acceleration should be the
same as for $M=0$. The  trajectory (\ref{traj2}) is similar to the
trajectory of a Galilean particle with mass $m$, and at first glance the
properly determined acceleration is the Galilean acceleration $a = F/m$.
However, the considered particles are essentially non-Galilean, since they
have positive and negative branches
of energy spectrum $E=\pm v  \Bigl({p_x}^2 + {p_y}^2\Bigr) $, which
correspond to the gapless energy spectrum with parabolic touching. That is
why instead of Galilean acceleraton $a = F/m$,  the characteristic
acceleration is determined by the  quantum fluctuation of $p_x$ on the
trajectory, as in the case of $M=0$, i.e. $a=F/M_{\rm eff} \sim v_F
F(1/mF)^{1/3}$.

So, in what follows the generalized acceleration in 2+1 anisotropic scaling
systems can be defined as
 \begin{eqnarray}
a = v_F F/|p_y|~~\mbox{for}~~  J\neq 2 ~,~p_y \neq 0 \,,
\label{acceleration_def1}
\\
 a = v_F F \Bigl(\frac{v}{F}\Bigr)^{1/(J+1)} = v_F T_U ~~\mbox{for}~~  J=2~,
~~\mbox{or~~for}~~  J>2~,~p_y=0\,.
\label{acceleration_def2}
\end{eqnarray}

\section{Accelerated observer interacting with quantum fluctuations}

In this section we consider the effect of the vacuum fluctuations on the
observer accelerated by the constant external force in the 2+1 quantum
vacuum with anisotropic scaling. As an example, we shall use the multilayer
graphene with rhombohedral (ABC...) stacking \cite{Katsbook}, in which the
topologically protected scaling exponent $z$ coincides with the number of
layers, $z=J$ \cite{Manes2007,KatsnelsonVolovikZubkov2012}.

\subsection{Trajectory of the detector, the interaction between the detector
and the fermionic excitations, and the click rate}

\label{sectclick}

In principle, one can use any trajectory of the detector. In this section we
shall consider for $J \ge 2$ the trajectory given by Eq. (\ref{traj0}). For
the special case $J = 1$ we consider different trajectory because Eq.
(\ref{traj0}) at $J=1$ would lead to the singularity.

In this section we assume that the detector interacts with the fermion
excitations via the term
\begin{equation}
V_{interaction} \sim {\psi}^+ O_A \psi  \label{VINT}
\end{equation}
 with some operators $O_A$. Different choices of these operators correspond
to different practical realizations of Unruh detector. In real graphene
various experimental situations may be suggested corresponding to different
practical realizations of the Unruh detector. In subsection \ref{experiment}
we shall briefly consider one of the realizations, - the Raman scattering
from the spot moving along the graphene sheet. In this case rough
approximation gives $V_{interaction}\sim \psi^+  \sigma^{a} \sigma^b \psi$ $
( a,b = 1,2)$ if the scattered light is not polarized and its direction is
orthogonal to the graphene sheet.
 It is worth mentioning that in this particular situation as well as in the
other possible experiments the microscopic description of the Unruh detector
is rather complicated. However, the properties of the spectra become
independent on the practical realization, and, in particular, on the form of
$V_{interaction}$ for the energies much larger than $T_U$, when the
semiclassical approximation works. Therefore, expressions obtained
semiclassically in subsection \ref{semiclas} are universal. Nevertheless,
for the completeness we consider several cases, when the problem can be
solved exactly. The considered cases correspond to idealized Unruh detectors
corresponding to different trajectories and different interaction terms Eq.
(\ref{VINT}).

The response function responsible for the transition of the detector to the
excited state is (Eq. (3) of \cite{unruhline}):
\begin{equation}
\dot{F}_{t}(\omega) = 2 {\rm Re} \, \int_0^{\infty} d f e^{-i\omega  f} \,
W(t, t -f)\label{W1}
\end{equation}

In Eq. (\ref{W1}) we encounter the Wightman function $W(t_1,t_2)$ of two
currents ${\psi}^+ O_A \psi$. However, for $t_1 > t_2$ it is equal to the
$T$-ordered two-particle Green function. We represent it as
\begin{eqnarray}
W(t_1, t_2) &=& \langle T {\psi}^+(t_1,x[t_1]) O_A \psi(t_1,x[t_1])
{\psi}^+(t_2,x[t_2]) O^+_A  \psi(t_2,x[t_2]) \rangle \nonumber\\&\sim &{\rm
Tr} \, O_A {\cal G}(t_1-t_2,x[t_1]-x[t_2])\, O^+_A {\cal
G}(t_2-t_1,x[t_2]-x[t_1]), \quad t_1>t_2 \label{Wightman}
\end{eqnarray}
where ${\cal G}$ is the single-fermion Green function. It is worth
mentioning that the expression for the click rate written in this form
resembles the conventional expression for the cross - section of the deep
inelastic processes like $e^+ + e^- \rightarrow {\rm hadrons}$. Both
quantities are related to the imaginary parts of the corresponding
polarization operators.  In fact, the derivations of these expressions are
similar. In Eq. (\ref{Wightman}) we should consider carefully the limit
$t_1-t_2 \rightarrow 0$ (see discussion at the end of this subsection).

The summation over $A$ is assumed in Eq. (\ref{Wightman}).
Below we concern several particular choices of $O_A$. In pure relativistic
$2+1$ theory there are the  following two natural choices:
$V_{interaction}\sim \bar{\psi} \psi = \psi^+ \sigma^3 \psi$ and
$V_{interaction}\sim \bar{\psi} \sigma^{\mu} \psi = \psi^+ \sigma^3
\sigma^{\mu} \psi, \mu = 1,2,3$.

In non - relativistic models of graphene the natural choices are:
$V_{interaction}\sim \bar{\psi} \psi = \psi^+ \sigma^3 \psi$,
$V_{interaction}\sim {\psi}^+  \psi$,  $V_{interaction}\sim \psi^+
\sigma^{a}  \psi, a = 1,2$, and  $V_{interaction}\sim \psi^+  \sigma^{a}
\sigma^b \psi, a,b = 1,2$. The latter case corresponds to Raman scattering
from the moving spot for the nonpolarized light incoming (and outgoing)
within the beam orthogonal to the graphene plane (see subsection
\ref{experiment}). In the relativistic case in $4D$ the Unruh detectors interacting with the
electron-hole pairs were considered in \cite{unruhfermi}.

The response function is expressed  as follows:
\begin{equation}
\dot{F}_{0}(\omega) \sim  {\rm Re} \, \int_0^{\infty} d t e^{-i\omega  t} \,
{\rm Tr} \, O_A {{\cal G}}(-t, -x[t]) O^+_A \, {{\cal
G}}(t,x[t])\label{click}
\end{equation}

We need to substitute into this expression the particular trajectory of the
detector and the particular operators $O_A$ corresponding to the interaction
of electron - hole pairs with the internal degrees of freedom of the
detector.
The integral in Eq. (\ref{click}) may be divergent at $t\rightarrow 0$. Let us
consider the polarization operator
\begin{equation}
\Pi(t,x)= i {\rm Tr} \, O_A {{\cal G}}(-t, -x) O^+_A \, {{\cal
G}}(t,x)\label{PO}
\end{equation}
The
expression $Q(\omega,x) = {\rm Im} \int_{0}^{+\infty} dt \,{\rm exp}
\Bigl( - i \omega t \Bigr) \,\Pi(t, x)$ gives the probability that the
detector at rest ($x=const$) absorbs ($\omega>0$) or emits ($\omega < 0$)
the electron - hole pair. Correspondingly,  $Q(\omega,x)$ should vanish for
$\omega>0$ because there are no free electron - hole pairs in vacuum. We
consider the analytical continuation of $\Pi(t,x)$ to negative values of
$t$, and extend integration to the whole real axis of $t$. Now we already
must remember, that we deal with the Wightman functions.  According to the
general properties of Wightman functions, $\Pi(-t,x) = - \Pi^*(t,x)$. We
close the contour in the lower half of the complex plane. The resulting
integral vanishes only if there are no poles within this contour. That is,
$\Pi(t, x)$ is analytical in the lower half of the complex $t$ - plane.

If the singularities appear at $t \in R$, we should go around them from
the bottom.
This gives the rules of going around poles. These rules are equivalent to
the existence of infinitely small imaginary contribution to $t$. We are to
substitute everywhere $\Pi(t-i \epsilon,x)$ instead of $\Pi(t,x)$.
Thus we shift in Eq. (\ref{click}) the integration to $t - i \epsilon$:
\begin{equation}
\dot{F}_{0}(\omega) \sim  {\rm Re} \, \int_0^{\infty} d t e^{-i\omega  t} \,
{\rm Tr} \, O_A {{\cal G}}(-t+i\epsilon, -x[t]) O^+_A \, {{\cal
G}}(t-i\epsilon,x[t])\label{click_reg}
\end{equation}
This is the standard way of the regularization of Wightman functions (see,
for example, \cite{BD,Akhmedov}). We adopt it throughout the text of the
present paper when the nonrelativistic results are concerned. However, for
the true relativistic case considered in section  \ref{J1} there is a
certain complication \cite{unruhline,unruhtime}.

\subsection{The two point Green function}

The two point Green function can be expressed as
\begin{equation}
{\cal G}(t,x) \sim \int d \epsilon d^2{ p} e^{-i \epsilon t+ i p_x x} {\cal
G}(\epsilon,p)
\end{equation}
Here ${\cal G}(\epsilon,p)$ is the fermion Green function in momentum space:
\begin{eqnarray}
{\cal G}(\epsilon, p) &=& \frac{i}{\epsilon  - H_0[p]} =  R[p]^+
\frac{i}{\epsilon - v |p|^J \sigma^3 } R[p],\nonumber \\
&&R[p]  =  \frac{1}{\sqrt{2}}\left(\begin{array}{cc} 1 &
\Bigl(\frac{p^*}{|p|}\Bigr)^J \\
-\Bigl(\frac{p}{|p|}\Bigr)^J & 1\end{array}\right), \quad p = p_1 + i p_2
\end{eqnarray}
In order to go around the poles correctly in the integral over $\epsilon$ we
consider the Wick rotated Green functions. We introduce the Euclidean time
$z$:  $t = -i z $. Integration over the Euclidean frequency $\omega$ ($\epsilon
= -i \omega$) gives for $z>0$:
\begin{eqnarray}
\int_{-\infty}^{\infty} d \omega \frac{e^{i \omega z}}{\omega - i v |p|^J
\sigma^3}
 =  2\pi i \left(\begin{array}{cc} e^{-v |p|^J z} & 0 \\
0 & 0\end{array}\right)
\end{eqnarray}

One can see, that for $z>0$ the Green function describes the propagation of
the excitation with positive energy (electron).
As a result
\begin{eqnarray}
{\cal G}(i z,x)& =&-\pi i \int d^2p\, e^{i (p, x) - v z |p|^J  }
\left(\begin{array}{cc} 1  &  \Bigl(\frac{p^*}{|p|}\Bigr)^J   \\
 \Bigl(\frac{p}{|p|}\Bigr)^J   & 1  \end{array}\right) \nonumber\\
&=&  \left(\begin{array}{cc} {\cal G}_{11} & {\cal G}_{12}  \\
{\cal G}_{12}  & {\cal G}_{11}  \end{array}\right)\nonumber\\
{\cal G}_{11}(iz,x) &=&- 2 \pi^2 i \int pdp\, J_0(|p|x)\, {\rm exp} (- v z
|p|^J)
\nonumber\\
{\cal G}_{12}(iz,x)  &= & - 2 \pi^2 i \int pdp\, i^{J}J_J(|p|x)\, {\rm exp}
(-v z |p|^J)
\end{eqnarray}
At $z<0$ we have
\begin{eqnarray}
\int_{-\infty}^{\infty} d \omega \frac{e^{i \omega z}}{\omega - i v |p|^J
\sigma^3}
 =  -2\pi i \left(\begin{array}{cc} 0 & 0 \\
0 & e^{-v |p|^J  |z|}\end{array}\right)
\end{eqnarray}
This corresponds to the propagation of the excitation with negative energy
(hole). For the elements of the Green function we have at $z<0$:

\begin{eqnarray}
{\cal G}_{11}(iz,x) &=& 2 \pi^2 i \int pdp\, J_0(|p|x)\, {\rm exp} (- v |z|
|p|^J)
\nonumber\\
{\cal G}_{12}(iz,x)  &= &  -2 \pi^2 i \int pdp\, i^{J}J_J(|p|x)\, {\rm exp}
(-v |z| |p|^J)
\end{eqnarray}
Next, we rotate back the Green function to real time $t = iz$, and get:
\begin{eqnarray}
{\cal G}_{11}(t,x) &=&- 2\, {\rm sign}(t) \, \pi^2 i  \int pdp\, J_0(|p|x)\,
{\rm exp} (-i v |t| |p|^J)
\nonumber\\
{\cal G}_{12}(t,x)  &= & - 2 \pi^2 i \int pdp\, i^{J}J_J(|p|x)\, {\rm exp}
(-i v |t| |p|^J)
\end{eqnarray}
 Below we shall describe several particular cases.

\subsection{The case $J=1$}
In this subsection we consider the (emergent) relativistic invariant
interaction term corresponding to $O_A = \sigma_3$ that gives
\begin{equation}
\dot{F}_{0}(\omega) \sim  2\, {\rm Re} \int_{0}^{\infty} d t e^{-i\omega t}
\, \frac{1}{(v^2 t^2 - x^2(t))^2},\label{FJ10}
\end{equation}
and the case $O_A = \sigma_a\sigma_b$ corresponding to Raman scattering. In
the latter case
\begin{equation}
\dot{F}_{0}(\omega) \sim   2\, {\rm Re} \int_{0}^{\infty}  d t e^{-i\omega
t} \, \frac{t^2}{(v^2 t^2 - x^2(t))^3}\label{FJ10R}
\end{equation}
Substituting here the trajectory Eq. (\ref{traj0}) with $x(0) = 0$ we get
the singularity. Therefore, we consider the motion along the following
trajectory, which experiences the thermal-like radiation
\cite{VolovikUnruh}. This is
\begin{equation}
x[t] = t u_0 {\rm th} \frac{at}{u_0} \label{trajJ1_}
\end{equation}
 with the limiting velocity which is smaller than the effective speed of
light, $u_0 < v$. The correlation function for the electron - hole pairs has
the poles at $t=0$ and at
\begin{equation}
t = \pm \frac{u_0}{a}[i\pi/2 \pm (1/2) {\rm  log} \frac{v+u_0}{v-u_0}]
\end{equation}
(compare this with Eqs. A.2, A.7 in \cite{VolovikUnruh}.) Therefore,   both
expressions for the click rate Eq. (\ref{FJ10}) and Eq. (\ref{FJ10R})
contain  Boltzmann exponent. Expressions Eq. (\ref{FJ10}), (\ref{FJ10R})
are divergent at $t \rightarrow 0$.
As it was explained at the end of section \ref{sectclick}, we apply the
conventional regularization of the Wightman functions  shifting the argument
$t$ to $t - i \epsilon$ with small $\epsilon$ (see also \cite{unruhline}).
In addition, the real part of $2 \int_0^{\infty}$ is equal to
$\int_{-\infty}^{\infty}$. The contour is closed either in the lower half of
the complex plane or in the upper half of the complex plane depending on the
sign of $\omega$.

For the case of the emission of the electron - hole pair the dominant nonexponential
term $\sim \omega^3$ is given by the pole at $t = 0$. For the
case of the absorbtion of the pair the pole at $t = 0$ does not enter the
expression for the click rate, and the term corresponding to the second pole
dominates. The corresponding tunneling exponent is given by
\begin{equation}
\dot{F}_{0}(\omega) \sim   {\rm exp}\Bigl({-\frac{\omega}{T}} \Bigr) , \quad
\omega>0\label{FJ1}
\end{equation}
with the temperature $T = \frac{a}{2\pi u_0}$. This is similar to Unruh
temperature  but with the limiting speed $u_0$ on the trajectory instead of
the effective "speed of light"  $v$.

\subsection{$J=1$: true relativistic case}
\label{J1}
In this subsection we consider the situation that cannot be realized in real
graphene. However, it resembles the true relativistic case and, therefore,
deserves the consideration. Namely, we suppose, that the internal degrees of
freedom of the detector are invariant under the Lorentz symmetry. This
occurs in true relativistic $2+1$ model.

We consider the hyperbolic motion
\begin{equation}
t(s) = \frac{v}{a} \, {\rm sh} \,( a s/v), \quad x(s) = \frac{v^2}{a}\, {\rm
ch}\, (as/v) - \frac{v^2}{a}
\end{equation}
with the linear acceleration $a$.
Instead of Eq. (\ref{W1}) we use
\begin{equation}
\dot{F}_{t}(\omega) = 2 {\rm Re} \, \int_0^{\infty} d S[f] e^{-i\omega
S[f]} \, W(t, t -f)\label{W10}
\end{equation}
where $S$ is the internal time of the detector, and
\begin{equation}
\dot{F}_{0}(\omega) \sim  {\rm Re} \, \int_0^{\infty} d S[t] e^{-i\omega
S[t]} \, {\rm Tr} \, {{\cal G}}(-t,-x) \, {{\cal G}}(t,x)\label{click0}
\end{equation}
instead of Eq. (\ref{click}).
We have
\begin{equation}
\dot{F}_{0}(\omega) \sim  {\rm Re} \, \int_0^{\infty} d s e^{-i\omega s} \,
\frac{1}{(v^2 t^2(s) - x^2(s))^2} \sim \int_{-\infty}^{\infty} d s
e^{-i\omega s} \, \frac{1}{{\rm sh}^4 \, \frac{a s}{2v}}\label{INTFJ1}
\end{equation}

The regularization described in section \ref{sectclick} applied here would
lead to the values of $\dot{F}_{t}(\omega)$ depending on $t$. Fortunately,
for $t=0$ the answer is obtained that coincides with the result obtained in Refs.
\cite{unruhline,Takagi1986},  when the finite small size of the detector is
taken into account. We suggest the following resolution of this puzzle.
Since the interaction between the detector and the electron - hole pairs in
this case is completely Lorentz invariant we {\it should} consider the
Wightman functions that describe emission/absorbtion in the reference frame
moving together with the detector. The arguments of section \ref{sectclick}
then lead us to the conclusion that the shift is to be applied to variable
$s$ (time in the moving reference frame) instead of $t$. Then the
regularized Wightman function $W(t[s_1-i\epsilon], t[s_2+i\epsilon])$
coincides with the Wightman function regularized as in Eq. (20) of
\cite{unruhline}.

Thus, we assume that the integration contour is shifted to $s - i \epsilon$
with small $\epsilon$.
Then the integral is equal to the sum over poles:
\begin{equation}
\dot{F}_{0}(\omega) \sim \left( \begin{array}{c} \omega^3 \sum_{k>0}
e^{-\frac{\omega}{T_U}k} \sim \frac{\omega^3}{e^{\frac{\omega}{T_U}}-1},
\quad \omega > 0\\
 \omega^3 \sum_{k\ge 0} e^{-\frac{|\omega|}{T_U}k} \sim
\frac{\omega^3 }{e^{\frac{|\omega|}{T_U}}-1} e^{\frac{|\omega|}{T_U}}, \quad
\omega < 0
\end{array}\right) \label{IJ1}
\end{equation}

Here $T_U = a/(2\pi v)$ is the Unruh temperature.  Thus the ratio of the
absorption probability to the emission probability is $\sim
e^{-\omega/T_U }$.

\subsection{The Dirac-Galilean case $J=2$,  the interaction term corresponds
to Raman scattering}

In this subsection we consider again the interaction term
$V_{interaction}\sim {\psi}^+ \sigma^a \sigma^b \psi$ that corresponds to
Raman scattering.
For the parabolic trajectory Eq. (\ref{traj2}) the problem can be solved
exactly.
Here
\begin{eqnarray}
{\cal G}_{11}(iz,x)&=&  - {\rm sign} (z) \, 2 \pi^2  \frac{i}{2 v  |z|}
e^{-\frac{1}{4  v |z|} x^2}
\nonumber\\
{\cal G}_{12}(iz,x)&=& - 2 \pi^2 \frac{i}{2v |z| x^2}(-4 v
|z|+e^{-\frac{1}{4  v |z|}x^2}x^2+4 e^{-\frac{1}{4  v |z|}x^2}v|z|)
\end{eqnarray}

and
\begin{eqnarray}
W(iz) &=&  {{\cal G}_{11}}(iz,x){{\cal G}_{11}}(-iz,-x) \sim
\frac{\pi^4}{v^2z^2} e^{-\frac{1}{2vz} x^2}
\end{eqnarray}

Next, we substitute $z =  i t$, and the trajectory $x[t]$ from Eq.
(\ref{traj2}) with $x[0]=0$:
\begin{eqnarray}
W(t) &\sim& -\frac{\pi^4 }{v^2t^2} e^{\frac{i}{2}v F^2 t^3}
\end{eqnarray}

The click rate (response function) has the form
\begin{eqnarray}
\dot{F}_{0}(\omega) &\sim&  \, - \int_{-\infty}^{\infty} d s e^{-i  \omega
s/T^{(2)}_U} \, \frac{1}{s^2} e^{\frac{i}{2} s^3} \label{FJ2}
\,.
\label{exact2}
\end{eqnarray}
Here $T^{(2)}_U$ is the Unruh temperature for vacua with $J=2$, it  is
determined by the generalized acceleration in Eq.(\ref{acceleration_def2})
and is given by Eq. (\ref{TU2}):
\begin{equation}
T^{(2)}_U= \Bigl(\frac{\hbar^2 F^2}{2m}\Bigr)^{1/3}
\label{Unruh2}
\end{equation}
(in this expression we have restored the Planck constant).

Expression Eq. (\ref{FJ2}) is linearly divergent at $s \rightarrow 0$.
The regularization described in section \ref{sectclick} effectively results
in the shift of the integration contour to $s - i \epsilon$ with small
$\epsilon$.
This is equivalent to the following regular expression for the click rate:
\begin{eqnarray}
\dot{F}_{0}(\omega) &\sim&  \, \pi (1-{\rm sign}(\omega))
\frac{|\omega|}{T_U^{(2)}} - \int_{-\infty}^{\infty} d s e^{-i  \omega
s/T^{(2)}_U} \, \Bigl(\frac{1}{s^2} e^{\frac{i}{2} s^3}-\frac{1}{s^2}\Bigr)
\label{FJ2__}
\,.
\label{exact2__}
\end{eqnarray}

In principle, the integral can be taken and expressed through the
generalized hypergeometric functions \cite{Bateman}. However, we do not
present here the corresponding lengthy result since the regular expression
of Eq. (\ref{FJ2__}) itself can be used effectively for the calculation of
the values of the distribution and for the drawing of the plot.
This distribution is highly asymmetric. For $\omega \rightarrow - \infty$ it
tends to infinity, while for $\omega \rightarrow + \infty$ it tends to zero.

Alternatively, this integral can be taken using the stationary phase
approximation. At $\omega >0$ and $\omega \gg T_U^{(2)}$ there are the
stationary points $s = \pm \sqrt{\frac{2 \omega}{3T_U^{(2)}}}$. The asymptotic expansion at
large $\omega$ may be obtained similar to that of the Airy function
\cite{LandauLifshitzV3}:
\begin{eqnarray}
\dot{F}_{0}(\omega) &\sim& \Bigl(\frac{3T_U^{(2)}}{ 2\omega}\Bigr)^{5/4} \,
{\rm cos}\Bigl[\Bigl(\frac{ 2\omega}{3T_U^{(2)}}\Bigr)^{3/2}\Bigr], \quad
\omega \gg T_U^{(2)}
\,,
\label{exact2_}
\end{eqnarray}
The derivation of this result admits the following interpretation.  If we
disregard the pseudospin structure, in the reference frame moving together
with the detector the one - particle hamiltonian has the form ${\cal H}  = -
2 v F p_x \tilde{t} + vp_x^2  + v p_y^2 $. It depends on time $\tilde{t}$, and the one - particle problem becomes non - stationary. In the above expressions
 the motion starts at $\tilde{t} = 0$ and ends at $\tilde{t} = t$.
The momentum $p$ does not depend on time but it depends on $t$ at the end of the time interval.
  The "classical" motion corresponds to the extremum of $\int_0^{t} {\cal H}(t) d\tilde{t} =  -v F p_x {t}^2 + vp_x^2 t + v p_y^2 t $ with respect to $p_x,p_y$, and gives
 $p_{cl,y}=0, p_{cl, x} = \frac{F}{2} t$. The corresponding
energy as a function of time satisfies $\int_0^{t} E(\tilde{t})d\tilde{t}  = - \frac{vF^2}{4} t^3$. In semiclassical
approximation the wave function of the electrons/holes contains
the factor ${\rm exp}( - i  \int_0^{t} E(\tilde{t})d\tilde{t}) \sim {\rm exp}( i \frac{vF^2}{4} t^3)$.
Because of the interaction with the detector (the interaction potential is
given by Eq. (\ref{VINT})) the hole may be transformed to electron at any
time. Therefore, the process is allowed, when first, the pair is created,
and second, it is annihilated. The amplitude  is proportional to ${\rm
exp}( - 2i \int_0^{t} E(\tilde{t})d\tilde{t}) \sim {\rm exp}(  i \frac{vF^2}{2} t^3)$. The amplitude of
the process with the two changes of the energy of the detector by $\pm
\omega$ is given by the Fourier transformation
\begin{equation}
{\cal A} \sim \int dt
(...)\, {\rm exp}( -i \omega t - 2 i \int_0^{t} E(\tilde{t})d\tilde{t}) = \int dt
(...)\, {\rm exp}( -i \omega t+ i \frac{vF^2}{2} t^3)\label{A2}
\end{equation}
 By unitarity the
total probability that the detector emits/absorbs the pair is proportional
to the imaginary part of $\cal A$.

Signs in Eq. (\ref{A2}) are important.
This is also important that positive values of $\omega$ correspond to the absorption of the electron -
hole pairs. Those signs can be checked as follows. Eq. (\ref{A2}) is similar to the quantum mechanical expression for the amplitude of the excitation of the quantum level with energy $2 E$ by the absorption of the light with frequency $-\omega > 0$. In the simplest case when $E$ is positive and does not depend on time the integral over $t$ gives the delta - function of $(\omega + 2 E)$. This delta - function may be nonzero for the excitation of the quantum level ($\omega < 0$). It vanishes for the inverse process ($\omega > 0$) because the system is assumed to be in the ground state.

At negative values of $\omega$ corresponding to the emission of pairs and at
$|\omega| \gg T_U^{(2)}$  the dominant term  corresponds to the Loran
expansion at $s=0$. This gives the power - like behavior:
\begin{eqnarray}
\dot{F}_{0}(\omega) &\sim& 2{\pi}\, \frac{|\omega|}{T_U^{(2)}}, \quad \omega
< 0.
\label{exact22_}
\end{eqnarray}

Equations (\ref{exact2}) and (\ref{exact2_}) are similar to equations which
determine thermal effects in systems at quantum criticality, where the
thermodynamic and kinetic parameters depend on frequency and temperature
through their ratio $\omega/T$. This generalizes the Hawking and Unruh
effects to the vacua with anisotropic scaling. This consideration clearly
demonstrates an intimate relation between Lorentz invariance of the field
theory and Boltzmann distribution in statistical physics. For the theory at $J=1$ with emergent Lorentz invariance
 the trajectory of the detector Eq. (\ref{traj}) being substituted to the expression for the click would give the singularity. Therefore, we considered the alternative trajectory Eq. (\ref{trajJ1_}) that leads to the
Boltzmann distribution of the pairs absorbed by the detector Eq. (\ref{FJ1}). The true Lorentz invariant theory leads to the Boltzmann distribution as well. For non-Lorentz
invariant theory with $J=2$, there is no Boltzmann distribution for absorbed particles. Instead, we have the law of Eq. (\ref{exact2_}).

\subsection{Arbitrary  $J \ge 2$, the semiclassical consideration}
\label{semiclas}

For arbitrary $J\ge 2$ we have (in the case of Raman scattering with $O_A =
\sigma_a \sigma_b$):
\begin{eqnarray}
\dot{F}_{0}(\omega) \sim \dot{\sigma}_{0}(\omega) &\sim & \int dt
e^{-i\omega t}{\cal G}_{11}(t,x){\cal G}_{11}(-t,-x)\nonumber\\ &= & 4\,
\pi^4  \int_{0-i0}^{\infty-i0} dt \int_0^{\infty} pdp
\int_0^{\infty}p^{\prime}dp^{\prime} \, J_0(p x[t])\, J_0(p^{\prime}
x[t])\,{\rm exp} (-i\omega t-i v t (p^J + [p^{\prime}]^J), \nonumber\\
&& x[t] = v F^{J-1} t^J
\end{eqnarray}
(For the definition of $\sigma(\omega)$ see subsection \ref{experiment}.)

Already for $J=3$ the elements of the two - point Green function are
represented as a rather complicated composition of generalized
hypergeometric functions. Therefore, for $J\ge 3$ we apply the semiclassical
approximation that works at $|\omega| \gg T_U$.
One of the advantages of the semiclassical consideration is that the
pseudospin degrees of freedom of the electrons and holes are neglected.
Therefore, the practical choice of operators $O_A$ entering the interaction
term Eq. (\ref{VINT}) does not matter.

Let us consider the trajectory Eq. (\ref{traj0}) for $J\ge 2$. The
dimensionality analysis demonstrates that the general form of the click rate
is
\begin{eqnarray}
\dot{F}_{0}(\omega) &\sim&   \int_{-\infty}^{\infty} d s e^{- i \omega
s/T_U^{(J)}} g_J(s)
\end{eqnarray}
with some function $g_J(t)$, and $T_U^{(J)}$  given by  Eq.(\ref{TU2}):
\begin{equation}
T_U^{(J)}=(F^J v)^{1/(J+1)}  = a/v_F \,.
\label{UnruhJ}
\end{equation}



At $\omega \gg T_U$ (absorption) in semiclassical approximation we have
\begin{eqnarray}
{\cal G}( t,x[t])& \sim &  \sum_{N=0,1;K=0..J-2}\frac{C_{NK}}{t^{J-1}}\,
{\rm exp}\Bigl(  i \, \frac{(J-1)}{J^{J/(J-1)}} \,  v F^J t^{J+1} e^{i \pi
\frac{2K+NJ}{J-1}}\Bigr),\label{CNK}
\end{eqnarray}
where $C_{NK}$ are constants, and
\begin{equation}
\dot{F}_{0}(\omega) \sim  {\rm Re} \, \int_0^{\infty} {d t} e^{- i \omega
t}{\cal G}^2( t,x[t])\label{INTF}
\end{equation}

  There are several stationary points in the complex plane. The dominant
contribution corresponds to the terms with $N=K=0$ in Eq. (\ref{CNK}).
Stationary phase approximation gives the following asymptotic behavior (the
consideration is similar to that of the Airy function
\cite{LandauLifshitzV3}):
\begin{eqnarray}
\dot{F}_{0}(\omega) &\sim&  \Bigl(\frac{T_U^{(J)}}{
\omega}\Bigr)^{\frac{5(J-1)}{2J}} \, {\rm cos} \Bigl[ \beta \, \Bigl(
\frac{\omega}{T_U^{(J)}}\Bigr)^{1+1/J}+{\rm const}\Bigr], \nonumber\\
&& \beta =  \frac{J^{J/(J-1)}}{[2(J^2-1)]^{1/J}(J+1)}
\label{semiunruh}
\end{eqnarray}

This result corresponds to the absorption of the electron - hole pairs by
the detector. Emission of the pairs by the detector corresponds to the
power - like behavior that can be extracted from the Loran expansion of the
integrand in Eq. (\ref{INTF}). Since the integration contour should be
shifted to $t - i0$ we can calculate the residue of the integrand at $t=0$.
The result gives the power - like dependence on $\omega$:

\begin{eqnarray}
\dot{F}_{0}(\omega) &\sim&   \Bigl(\frac{|\omega|}{T_U^{(J)}}\Bigr)^{2J-3},
\quad |\omega| >> T^{(J)}_U, \quad \omega < 0.
\label{exactJ22_}
\end{eqnarray}

Applying the same technique to Eq. (\ref{INTFJ1}) we can obtain the dominant
power - like term for the emission of the pairs in the relativistic case
$\sim |\omega|^3$ that can also be seen in the exact result Eq. (\ref{IJ1}).
This behavior is in some sense similar to that of considered in \cite{BH}
and in   Sec. 31.4 of \cite{Volovik2003}, where
the power-law asymptote for radiation from rotating black hole is obtained
by tunneling method.

\subsection{Discussion of possible experiments with the accelerated detector
of electron - hole pairs}
\label{experiment}

 Experiments for the case of ``accelerating detector'' look difficult but
not impossible: one should create a laser spot moving along the graphene
surface with acceleration and measure the Raman scattering from such a spot.
The emission (adsorption) of electron-hole pairs by the moving spot
corresponds to Stokes (anti-Stokes) Raman scattering, with the lower
(higher) frequency of the scattered light in comparison with the incident
one. In order to check the distributions obtained in this section this spot
should move along the corresponding trajectories given by Eq. (\ref{traj0}).
{One should expect a huge asymmetry, with the oscillating factor for the dependence on $\omega$
 for the absorption versus the power-law  for the emission}.
Unruh effect corresponds to the absorption of the electron - hole pairs by the
accelerated detector, or, better to say, the Unruh effect probability is the
ratio of the absorption probability to the emission probability. In order to
model the Unruh detector the absorbtion of the incoming photon and emission
of the secondary photons should occur in the small vicinity of the spot and
without the essential delay. Otherwise the spot cannot be considered as the
point - like detector moving along the given trajectory.

 The detailed consideration of this process, and the questions related to
the delays and the correlation length of Raman scattering from the spot is
out of the scope of the present paper. Nevertheless, in order to understand,
what plays the role of the operator $O_A$ of Eq. (\ref{VINT}) in this case,
let us consider the simplest diagram that describes the Raman scattering.
The incoming photon is transformed to the electron - hole pair. Next, the
electron and the hole scatter elastically on the atoms of the crystal
lattice. During this scattering process the momentum of the incoming photon
(minus the momentum of the outgoing photon) is absorbed by the crystal
lattice.   After that the scattered electron emits the outgoing
photon\footnote{Actually, the hole may emit the photon, and this may happen
before the scattering on the crystal lattice. Also, the photon may first
scatter elastically on the lattice, and then be transformed to the
electron - hole pair. But the description of the corresponding processes
will lead to the same final answer for the operator $O_A$.}. The initial
state is vacuum, the final state includes the created electron - hole  pair.

For the cross section of the Raman scattering from the moving spot we have
\begin{eqnarray}
\sigma_T(\omega) &\sim  &
\int_{-\infty}^T dt \int_{-\infty}^{T} dt^{\prime} e^{-i \omega (t -
t^{\prime})} \sum_{a,b} \langle 0| \psi^+(x[t^{\prime}])  \sigma_b \sigma_a
\psi(x[t^{\prime}]) \psi^+(x[t])  \sigma_a \sigma_b \psi(x[t])|0 \rangle
\label{RS2_}
 \end{eqnarray}
Here the sum is over the polarizations $a,b = 1,2$ of incoming and outgoing
photons.
The so - called click rate \cite{unruhline} is given by the derivative of
$\sigma$ with respect to $T$ in the integral over $t$:
\begin{eqnarray}
\dot{\sigma}_{0}(\omega) &\sim  & 2\, {\rm Re}\,
\int_{-\infty}^{0} dt  e^{-i \omega t} \sum_{a,b} \langle 0| \psi^+(x[t])
\sigma_b \sigma_a \psi(x[t]) \psi^+(x[0])  \sigma_a \sigma_b \psi(x[0])|0
\rangle \label{RS2__}
\end{eqnarray}
We come to Eq. (\ref{W1}) calculated for the interaction term Eq.
(\ref{VINT}) with $O_A = \sigma_a\sigma_b$.  Eq. (\ref{RS2__}) was derived
for the emission of electron - hole pairs. This corresponds to negative
values of $\omega$. For the spot at rest $\dot{\sigma}(\omega)$ vanishes for
$\omega > 0$ since there are no free electron - hole pairs in vacuum.
However, for the accelerated spot $\dot{\sigma}(\omega)$ already does not
vanish for positive $\omega$. This means that the energy of the scattered
photon may be larger than the energy of the incoming photon. This can be
interpreted as the detection of the electron - hole pairs by the  Unruh
detector.

 It is worthwhile to note that the moving laser beams were successfully used
to study quantum phenomena in ultracold gases \cite{Miller2007}. The laser
spot moving along the circular trajectory in BEC has been suggested
\cite{Takeuchi2008} to study the analog of Zel'dovich-Starobinsky effect
\cite{BH}, which is similar to the effects discussed here.

\section{The system in the presence of electric field}

In the presence of external electric field electron - hole pairs are created
due to the Schwinger mechanism. These pairs may be detected in some way by
the detector at rest. This situation is somehow similar to that of the
accelerated detector considered in the previous section. Namely, free
electrons and holes would move with the acceleration (in opposite
directions). In some sense we may speak on the ``accelerated vacuum'' in
this situation.

\subsection{Distribution of electron-hole pairs as a function of energy}

The one-particle Hamiltonian for  $2D$  massive fermions with anisotropic
scaling
$E^2=v^2(p_x^2+p_y^2)^{J}$ in electric field ${\cal E}$ has the form
\begin{equation}
H=\left(
\begin{array}{cc}
{\cal E}x & v\Bigl(\hat{p}_{x}-i\hat{p}_y\Bigr)^{J} \\
v\Bigl(\hat{p}_{x}+i\hat{p}_y\Bigr)^{J} & {\cal E}x%
\end{array}%
\right)   \label{eq1}
\end{equation}%
Here $\hat{p}_{j}=-i\partial _{j}$; we  use the units $\hbar =e=1$.

Integration over the classically forbidden region   gives  us the pair
production probability (probability of the Schwinger effect)
\cite{KatsnelsonVolovikZubkov2012}
\begin{eqnarray}
|\eta _{0}|^{2} &=&\exp{\left(- \alpha \frac{v}{\cal E} p_y^{J+1}\right)},
\nonumber \\ && \alpha = 2 B\Bigl(\frac{1}{2},\frac{J+2}{2}\Bigr)
\label{PairProduction}
\end{eqnarray}%

The energy of the electron-hole pair is $\omega = 2 v (p_x^2+p_y^2)^{J/2}$.
The distribution of the created pairs as a function of energy has the form:
 \begin{eqnarray}
f(\omega) &=& \int  dp_x  dp_y e^{-\alpha \frac{v}{\cal E}
p_y^{J+1}}\delta(\omega - 2 v (p_x^2+p_y^2)^{J/2} )\nonumber\\
&\sim &
\int_{-\Bigl(\frac{\omega}{2T}\Bigr)^{1/J}}^{\Bigl(\frac{\omega}{2T}\Bigr)^{
1/J}} dq \frac{{\rm exp}\Bigl(-\alpha \Bigl(
\Bigl(\frac{\omega}{2T}\Bigr)^{2/J} -
q^2\Bigr)^{(J+1)/2}\Bigr)}{\omega^{1-2/J}\Bigl(
\Bigl(\frac{\omega}{2T}\Bigr)^{2/J} - q^2\Bigr)^{1/2}}
\nonumber\\
&\sim & \int_{-1}^{1} dx \frac{{\rm exp}\Bigl(-\alpha
\Bigl(\frac{\omega}{2T}\Bigr)^{(J+1)/J} \Bigl(1 -
x^2\Bigr)^{(J+1)/2}\Bigr)}{\omega^{1-1/J} \Bigl( 1 - x^2\Bigr)^{1/2}}
\label{PairProductionE}
\nonumber\\
&\sim &\frac{1}{\omega^{1-1/J}} \int_0^{1} dz \exp\Bigl(-\alpha
\Bigl(\frac{\omega}{2T}\Bigr)^{(J+1)/J} z \Bigr) z^{1/(J+1) - 1}\Bigl( 1 -
z^{2/(J+1)}\Bigr)^{1/2 - 1}
\label{PairProductionE}
\end{eqnarray}%

Let us introduce the Unruh temperature $T = T_U^{(J)}$ in correspondence
with Eq. (\ref{TU2}):
\begin{equation}
T_U^{(J)}=({\cal E}^J v)^{1/(J+1)}
\end{equation}
(we use here ${\cal E}$ instead of $F$.)

\subsection{Particular cases}

For $J=1$ we arrive at
\begin{eqnarray}
f(\omega) &\sim & \int_0^{1} dz \exp^{-\alpha
\Bigl(\frac{\omega}{2T}\Bigr)^{2} z } z^{1/2 - 1}\Bigl( 1 - z \Bigr)^{1/2 -
1}\nonumber\\
&=& \pi \exp\Bigl(-\frac{\alpha}{2} \Bigl(\frac{\omega}{2T}\Bigr)^{2}\Bigr)
I_0 \left(\frac{\alpha}{2} \Bigl(\frac{\omega}{2T}\Bigr)^{2}\right)
\end{eqnarray}

This distribution was also derived in Ref. \cite{KLR2010}, where the
spontaneous recombination of electrons and holes created due to the
Schwinger effect in monolayer graphene was considered.

For $J\ge 2$ the function $f(\omega)$ is expressed through the generalized
hypergeometric functions (for their definition see \cite{Bateman}). In
general case $\omega^2 f(\omega)$ is the analytical function of $\zeta=
\alpha \Bigl(\frac{\omega}{2T}\Bigr)^{(J+1)/J} $:

\begin{equation}
\omega^2 f(\omega) \sim
\zeta \int_0^{1} dz \exp\Bigl(-\zeta z \Bigr) z^{1/(J+1) - 1}\Bigl( 1 -
z^{2/(J+1)}\Bigr)^{1/2 - 1}\label{intrep}
\end{equation}

This integral is convergent for any values of $J$. It can be represented as
a composition of the generalized hypergeometric functions. For example, for
$J = 2$:
\begin{equation}
\omega^2 f(\omega) \sim  -\frac{2}{3}\zeta^2\, _3F_4\Bigl[
\begin{array}{cc}
\frac{2}{3}, 1, \frac{4}{3};& \frac{\zeta^2}{4}\\
 \frac{5}{6}, \frac{7}{6}, \frac{3}{2}, \frac{3}{2} &   \\
\end{array}\Bigr]
+\frac{ \pi}{2}\zeta \, _2F_3\Bigl[\begin{array}{cc}
\frac{1}{6}, \frac{5}{6};& \frac{\zeta^2}{4}\\
 \frac{1}{3}, \frac{2}{3}, 1& \\
\end{array}\Bigr]
\end{equation}

In this expression the functions $_pF_q$ at small values of the argument are
given by \cite{Bateman}
\begin{equation}
_pF_q\Bigl[
\begin{array}{cc}
a_1, ..., a_p;& z\\
 b_1, ... , b_q&   \\
\end{array}\Bigr] = \sum_{n=0}^{\infty} \frac{(a_1)_n ... (a_p)_n}{(b_1)_n
... (b_q)_n} \, \frac{z^n}{n!}, \quad (a)_n = a (a+1)...(a+n-1) \label{GHF}
\end{equation}

\subsection{Asymptotic expansions}

The expansion of Eq. (\ref{GHF}) works in a vicinity of $z = 0$ and does not
work for $z \rightarrow \infty$.
In order to evaluate the distributions at $\zeta \rightarrow \infty$ we use
integral representation Eq. (\ref{intrep}). We obtain:
\begin{eqnarray}
f(\omega) &\sim & \zeta^{\frac{1-J}{J+1}}  \Gamma(1/(J+1))
\frac{1}{\zeta^{1/(J+1)}}\nonumber\\
&\sim & \frac{2 T_U}{ \omega} , \quad \omega \gg T_U \label{PLF}
\end{eqnarray}

One can see that the dominant term is power - like. Among the other terms
there are also the subdominant terms that contain the tunneling exponent
$e^{-\zeta}$ (this may be obtained as an expansion near $z=1$ in the
integral of Eq. (\ref{intrep})).
 It is instructive to compare this with the behavior of the distributions of
electron - hole pairs detected/emitted by the moving Unruh detector.   Eqs.
(\ref{intrep}),(\ref{PLF}) are to be compared with Eq (\ref{semiunruh}). In
both cases there is the dependence on $\zeta =
\Bigl(\frac{\omega}{2T^{(J)}_U}\Bigr)^{1/J+1}$. This indicates that the two
phenomena may  indeed be similar, but there is no direct correspondence.  In
the distribution of the electron - hole pairs created in external Electric
field the leading term is power - like. This is somewhat similar to the
distribution of pairs absorbed by the accelerated detector Eq.
(\ref{semiunruh}). However, the powers are different, and in the latter case
the distribution contains also the oscillating factor.

\subsection{ $f(\omega)$ expressed through the Green functions}

 Expression Eq. (\ref{PairProductionE}) can be rewritten as
\begin{eqnarray}
f(\omega) &=& \int d\tau e^{i\omega \tau} \int  dp_x dp_y e^{-\alpha
\frac{v}{\cal E} p_y^{J+1}-i \tau 2 v (p_x^2+p_y^2)^{J/2} }\sim  \int d\tau
e^{i\omega \tau} W(\tau) \label{fw}
\end{eqnarray}

Recall that the expression $Q = - \alpha \frac{v}{\cal E} p_y^{J+1}$ has its
origin at
\begin{equation}
Q = 2\, {\rm Im}\, \int {\cal P} dx,
\end{equation}
where
\begin{equation}
{\cal P} = \sqrt{\Bigl((E-{\cal E} x)/v\Bigr)^{2/J}-p_y^2}
\end{equation}
is the classical momentum.   This, in turn,  may be obtained as a
semi-classical approximation to the representation
\begin{eqnarray}
f(\omega) &=&{\rm Re}\, \int d\tau e^{i\omega \tau} \langle \bar{\psi}(0,0)
\psi(0,0) \bar{\psi}(\tau,x)  \psi(\tau,x) \rangle dx \nonumber\\
&\sim & {\rm Re} \int d\tau e^{i\omega \tau} {\rm Tr} \,  {\cal G}_{\cal
E}(\tau,x)\,  {\cal G}_{\cal E}(-\tau,-x)dx
\end{eqnarray}
Here fermion Green functions in external electric field are present.
This expression is to be compared with Eq. (\ref{click}).

\subsection{Discussion of possible experiments with the ``accelerated
vacuum'' of graphene}

 Relevant energies of electron excitations for this case lie in the
infrared, which allows us to suggest for the detection an efficient optical
tool, namely, the electron Raman scattering (see, e.g., \cite{Raman}).  One
should probe the dependence of the Raman spectra on electric field. This
will correspond to our case of ``accelerating vacuum''. Another way to probe
the distribution of the electron - hole pairs on energy is to consider the
spontaneous recombination of electrons and holes forming the electron - hole
plasma \cite{KLR2010} which may be in principle probed via luminescence.

 Of course, estimation of the Unruh temperature is crucially important for
the choice of the proper experimental technique. First, notice that graphene
is very robust material with respect to electric breakdown; the current
densities such as $(3-4) \times 10^{4}$ A/m are reachable (for recent
references, see \cite{Barreiro1,Barreiro2}) which corresponds to the
electric fields ${\cal E}_c \approx 10^8 - 10^9$ V/m. According to Eq.
(\ref{Unruh2}), for the case of bilayer graphene it corresponds to Unruh
temperature as large as 0.5 eV $\approx$ 6000 K. So strong electric fields
produce a strong heating \cite{Barreiro2} and this, of course, can be a
serious problem but even for the orders of magnitude weaker fields the Unruh
temperature can be still quite high (it is about 10 meV for ${\cal E}
\approx 3 \times 10^{6}$ V/m). Relevant energies of electron excitations for
this case lie in the infrared, which allows us to suggest for the detection
an efficient optical tool, namely, the electron Raman scattering (see, e.g.,
\cite{Raman}).

One should keep in mind, however, that below the energy of the order of 10
meV the effects of trigonal warping and electron-electron interactions
become important \cite{Katsbook,mayorov} with the reconstruction of
parabolic touching point to several conical touching points at low energies.
This means that for the electric fields ${\cal E} \approx \ll 10^{6}$ V/m)
we will probe these conical points rather than the parabolic point, and the
Unruh spectra should be similar to those for the single-layer graphene.
However, the Fermi velocity for these new conical points is much smaller
than in the single-layer graphene ($1.4 \times 10^5$ m/s and $9 \times 10^5$
m/s, respectively \cite{mayorov}) which will lead to an essential increase
of the Unruh temperature (inversely proportional to the Fermi velocity), the
effect which looks very interesting by itself.

Alternative types of detection of the created electron-hole pairs and their
spectroscopy are emission M\"{o}ssbauer spectroscopy (one should put on
graphene, e.g., radioactive $^{57}$Co) \cite{Wertheim} or electron spin
resonance \cite{Slichter} of magnetic adatoms on graphene. In both cases the
spectra are modulated by the fluctuating electron density.  However, optical
methods (Raman spectroscopy) look the most promising and realistic.

\section{Massive 1D fermions}

In this section we consider the dynamics of the one - dimensional system
with the Hamiltonian given by Eq. (\ref{H1}), where $p_y$ is fixed and is
equal to $M$ playing the role of effective mass of the fermionic
excitations. The multilayer graphene itself (at ABC stacking) can be
considered as a collection of such $1D$ systems corresponding to different
values of $M$.
The one - dimensional system considered in this section is naturally
realized in graphene nanoribbons that is the thin strip of the multilayer
graphene. The values  $p_y$ are quantized there \cite{Katsbook}, and the
subsystems with different values of $M=p_y$ can be distinguished from each
other (see, e.g., the discussion of conductance quantization in Section 5.5
of the book \cite{Katsbook}).

\subsection{ M\"uller temperature for 1+1 massive fermions in electric
field}

In the presence of constant external electric field we deal with the
``accelerated vacuum''.
 We consider the Hamiltonian Eq. (\ref{H1}) in the case when $p_y$ is fixed
and equal to $M$.  Then $p = p_x +i M$, and one-particle Hamiltonian for  1D
massive fermions with anisotropic scaling
$E^2=v^2(p^2+M^2)^{J/2}$ in electric field ${\cal E}$ has the form
\begin{equation}
H=\left(
\begin{array}{cc}
{\cal E}x & v\Bigl(\hat{p}_{x}-iM\Bigr)^{J} \\
v\Bigl(\hat{p}_{x}+iM\Bigr)^{J} & {\cal E}x%
\end{array}%
\right)   \label{eq1}
\end{equation}%

Integration over the classically forbidden region   gives  us the pair
production probability
\cite{KatsnelsonVolovikZubkov2012}
\begin{eqnarray}
|\eta _{0}|^{2} &=&e^{-2JB\Bigl(\frac{3}{2},\frac{J}{2}\Bigr)vM^{J+1}/{\cal
E}},   \,.
\label{PairProduction}
\end{eqnarray}%
With the definition of generalized acceleration $a=v_F{\cal E}/M$ of
Eq.(\ref{acceleration_def1}) this expression resembles the thermal
distribution. Namely, we may represent
\begin{eqnarray}
|\eta _{0}|^{2} &=&\exp\left(- \frac{v M^J}{T_{\rm M}}\right),  \nonumber \\
&&T_{\rm M} =\gamma^{-1}\hbar  a/v_F~~,~~a=v_F\frac{\dot p}{M}=
v_F\frac{\cal E}{M}
~~,~~\gamma=\frac{2J}{\pi}B\Bigl(\frac{3}{2},\frac{J}{2}\Bigr)\,.
\label{PairProduction4}
\end{eqnarray}%
with the analogue of  M\"uller temperature \cite{Muller1977} $T_M$.
(In the last expression we restore Plank constant $\hbar$.)
In the relativistic systems ($J=1$) the M\"uller temperature $T_{\rm
M}=\hbar a/\pi c$ is twice the Unruh temperature  $T_{\rm U}=\hbar a/2\pi c$
\cite{Unruh1976}, see e.g. recent paper  \cite{LabunRafelski2012}.
 The only difference from the relativistic system is that in anisotropic
scaling system the pre-factor  depends on $J$.

\subsection{Unruh effect at $|\omega| \gg T_U$}

In this subsection we consider the situation, when the detector moves in the
given $1D$ system along the trajectory given by Eq. (\ref{traj}). With  $p =
p_x +i M$ one has the Green's function of $1D$ quasiparticles
\begin{eqnarray}
\tilde{\cal G}(i t,0)& =&2 \pi i \int d p_x\, e^{i p_x x - v |t| |p|^J  }
\left(\begin{array}{cc} 1  &  {\rm sign} (t) \,
\Bigl(\frac{p^*}{|p|}\Bigr)^J   \\
{\rm sign} (t) \,  \Bigl(\frac{p}{|p|}\Bigr)^J   & 1  \end{array}\right)
\nonumber\\
&=&  \left(\begin{array}{cc} G_{11}(it) & G_{12}(it)  \\
G_{21}(it)  & G_{11} (it) \end{array}\right)\nonumber\\
G_{11}(it) &=& (2 \pi) i \int dp_x\,  {\rm exp} (i p_x x - v |t|
(p_x^2+M^2)^{J/2})
\nonumber\\
G_{12}(it)  &= & {\rm sign} (t) \,  (2 \pi) i \int dp_x \,  {\rm exp} (ip_x
x-v |t| (p_x^2+M^2)^{J/2})   \Bigl(\frac{p_x-iM}{(p_x^2+M^2)^{1/2}}\Bigr)^J
\nonumber\\
G_{21}(it)  &= &{\rm sign} (t) \,   (2 \pi) i \int dp_x \,  {\rm exp} (ip_x
x-v |t| (p_x^2+M^2)^{J/2})   \Bigl(\frac{p_x+iM}{(p_x^2+M^2)^{1/2}}\Bigr)^J
\end{eqnarray}

We evaluate the Green function in the semi - classical approximation. Then
\begin{eqnarray}
{\cal G}( t,x[t])& \sim & \int d p_x\, e^{i p_x x - i v |t| |p|^J  }
\end{eqnarray}

Stationary phase conditions give
\begin{equation}
x = v |t| J p_x (p_x^2+M^2)^{J/2-1}
\end{equation}

Below we consider the two particular cases ($J=1,2$) of Unruh radiation of
massive particles with the energy near to the threshold $2E(p=0)=2vM^J$.

\begin{enumerate}

\item{$J=1$.}

In this case we have
\begin{eqnarray}
{\cal G}( t,x[t])& \sim & \frac{e^{i\pi/4}}{v|t|^{1/2}} {\rm exp}\Bigl(  -i
v M \sqrt{1-\Bigl(\frac{x}{v t}\Bigr)^2}|t|\Bigr)
\end{eqnarray}
As a result
\begin{equation}
\dot{F}_{0}(\omega) \sim  {\rm Re} \int_{0}^{\infty} d t \frac{i}{v t} {\rm
exp}\Bigl(-i\omega t - 2 i v M \sqrt{1-\Bigl(\frac{x}{v t}\Bigr)^2}
t\Bigr)\sim  \int_{-\infty}^{\infty} d t \frac{i}{v t} {\rm
exp}\Bigl(-i\omega t - 2 i v M \sqrt{1-\Bigl(\frac{x}{v t}\Bigr)^2} t\Bigr)
\end{equation}

Here the trajectory of the detector $x(t)$ is given by Eq. (\ref{traj}):
\begin{equation}
x(t) = v \Bigl({t}^2 + \frac{v^2}{a^2}\Bigr)^{1/2} - v
\frac{v}{a}\label{traj1}
\end{equation}

We also use the following parametrization of this trajectory:
\begin{equation}
t(s) = \frac{v}{a} \, {\rm sh} \, (a s/v), \quad x(s) = \frac{v^2}{a}\, {\rm
ch}\, (as/v) - \frac{v^2}{a}
\end{equation}

We come to
\begin{equation}
\dot{F}_{0}(\omega) \sim   i \, \int_{-\infty}^{\infty} d s \frac{1}{{\rm
th}\, (as/v)} {\rm exp}\Bigl(-\frac{i v}{a}\Bigl[\omega \, {\rm sh} \, (a
s/v) + 4  v M\, {\rm sh} \frac{a s}{2v}\Bigr]\Bigr)
\end{equation}

At $\omega \sim 2v M \gg T_U = \frac{a^2}{v^3 M} $ we can apply the
stationary phase approximation. For $0<\omega < 2 v M$ and $\frac{|\omega -
2 vM|}{2vM}<<1$ there are no stationary points at real $s$ but there is the
stationary point at imaginary values of $s$ in the lower half of the complex
plane. Therefore, the absorbtion of the electron - hole pairs is described
by the tunneling exponent.
One can define the effective chemical potential $\mu_{eff} = 2  v M$ and the
effective Unruh temperature $T_U = \frac{a^2}{v^3 M} = \frac{F^2}{v  M^3}$.
Then
\begin{equation}
\dot{F}_{0}(\omega)  \sim   e^{- 4 \sqrt{6}  \Bigl(\frac{|\omega -
\mu_{eff}|}{T_U}\Bigr)^{1/2}}, \quad  |\omega - 2 v M| \ll 2vM, \quad \omega
\gg a/v, \quad \omega < 2 v M
\end{equation}
It is worth mentioning that in this case the Unruh temperature is not given
by Eq. (\ref{TU2}). Instead we have the relation that involves both mass
parameter $M$ and the acceleration $a$.
For $\omega > 2 v M$ there is the stationary point for the real value of $s$
that results in the appearance of the oscillating factor in
$\dot{F}_0(\omega)$:
\begin{equation}
\dot{F}_{0}(\omega)  \sim   {\rm sin}\Bigl(4 \sqrt{6}  \Bigl(\frac{|\omega -
\mu_{eff}|}{T_U}\Bigr)^{1/2}\Bigr), \quad  |\omega - 2 v M| \ll 2vM, \quad
\omega \gg a/v, \quad \omega > 2 v M
\end{equation}

\item{$J=2$.}

In this case
\begin{eqnarray}
{\cal G}( t,x[t])& \sim & \frac{e^{i\pi/4}}{|t|^{1/2}} {\rm exp}\Bigl(
\frac{i}{4} v |t| ( F t - M) ( F t + M) \Bigr)
\end{eqnarray}
As a result
\begin{equation}
\dot{F}_{0}(\omega) \sim   \int_{-\infty}^{\infty} d t \frac{i}{v t}{\rm
exp}\Bigl(-i\omega t + \frac{i}{2} v t ( F t - M) ( F t + M) \Bigr)
\end{equation}

Next, we use the stationary phase approximation. It is valid  at $ |\omega|
\gg T_U = (v F^2)^{1/3}$. Positive values of $\omega$  correspond to the
absorbtion of the electron - hole pairs.  The Unruh temperature for this
case  $T_U \sim (v F^2)^{1/3}$ is equal to that of defined by Eq.
(\ref{TU2}) with the generalized acceleration of
Eq.(\ref{acceleration_def2}) as expected for the $J=2$ case, when the
generalized acceleration does not depend on $M$. We denote $\mu_{eff} = 2v
M^2$, and get similar to the case of the Airy function
\cite{LandauLifshitzV3}:
\begin{equation}
\dot{F}_{0}(\omega)  \sim  {\rm sin}\Bigl(\frac{2 \sqrt{6}}{9 }
\Bigl(\frac{\omega+\mu_{eff}}{T_U}\Bigr)^{3/2}\Bigr), \quad \omega \gg T_U,
\quad \frac{|\omega - 2 v M^2|}{2 vM^2} \ll 1
\end{equation}

\end{enumerate}

\subsection{Suggested experiments}

To probe the distributions obtained in this section we suggest the same
methods as suggested in Sections III and IV but for the graphene
nanoribbons. Namely, Raman scattering and spontaneous recombination can be
used to probe the spectrum of electrons and holes created during the
Schwinger process. The Raman scattering from the moving spot can be used to
probe the Unruh effect itself.

\section{Conclusions}

Vacua with anisotropic scaling, such as ${\bf r} \rightarrow b {\bf r}$ , $t
\rightarrow b^z t$,
represent new topological classes of quantum vacua, where the scale
parameter $z$ is determined
by the topological charge of the quantum vacuum. The relativistic invariance
emerges only in one of the the universality classes of vacua, where topology
supports the isotropic scaling $z=1$.
Anisotropic scaling naturally emerges in many condensed matter systems,
including multilayer graphene. That is why it is instructive to consider
general properties of such quantum vacua.
Here we discussed the quantum effects, such as Unruh radiation and Schwinger
pair production, in vacua which belong to universality classes with
topologically protected  anisotropic scaling.  Though  these effects do not
possess thermal properties relevant for the relativistic class of quantum
vacua with $z=1$, some features are proved to be universal. Both effects,
Unruh radiation and Schwinger  pair production, are characterized by the
modification of correspondingly Unruh and M\"uller temperatures.
This characteristic temperature  is proportional to the acceleration
properly generalized to systems with the anisotropic scaling in Eqs.
(\ref{acceleration_def1})
and (\ref{acceleration_def2}). As a rule such temperature does not describe
any
thermal distribution of radiated excitations. Instead it acts as the
temperature, which enters the thermodynamic and kinetic parameters of the
many-body system in the vicinity of the quantum phase transition, where the
vacuum experiences the properties of quantum criticality.

\section*{Acknowledgements}
M.A.Z. kindly acknowledges useful discussions with M.Lewkowicz,
B.Rosenstein, and E.Akhmedov. This work was partly supported by RFBR grant
11-02-01227, by Grant for
Leading Scientific Schools 6260.2010.2, by the Federal Special-Purpose
Programme 'Cadres' of the Russian Ministry of Science and Education. MIK
acknowledges a
financial support by FOM (the Netherlands). GEV acknowledges financial
support  by the EU 7th Framework Programme
(FP7/2007-2013, grant $\#$228464 Microkelvin) and by the Academy of
Finland through its LTQ CoE grant (project $\#$250280).

\end{document}